# Fully automatic fabrication of fibre Bragg gratings using an AI-powered femtosecond laser inscription system

Wenbo Liu[#], Guiyuan Cao[#], Zian Liu, Hongyang Chen, Hao Zhang, Renjie Li, Keng-Te Lin, Han Lin, Baohua Jia

**Abstract**
Fibre Bragg gratings (FBGs) are widely used in optical sensing and communication systems. Femtosecond laser inscription (FLI) enables hydrogen-free, thermally stable, high-resolution, and complex structures of FBG fabrication, but its practical application is limited by manual operation, low throughput, and sensitivity to laser alignment. In this study, we present an AI-powered FLI system that enables automated, stable, and efficient FBG fabrication. By integrating a Multi-Layer Perceptron (MLP) model for real-time fabrication position correction, the system maintains precise laser alignment (-0.6 to 0.2 µm of the fibre core plane) and ensures consistent processing. Strong and weak FBGs were fabricated in different types of fibres, and their spectral characteristics—including central wavelength, reflectivity, and FWHM—exhibited high stability and repeatability. The results demonstrate that the proposed AI-powered FLI system significantly reduces manual intervention while achieving reliable FBG performance. This approach holds great promise for scalable, high-throughput FBG production and can be extended to the fabrication of arbitrary FBG structures across various fibre types. With further training and model refinement, the AI-powered FLI provides a scalable and intelligent platform for next-generation automated FBG manufacturing.

## 1. Introduction

A fibre Bragg grating (FBG) is a periodic modulation of the refractive index (RI) along the core of an optical fibre, which reflects light at a specific wavelength—known as the Bragg wavelength—while transmitting all others[1, 2]. Featuring high precision, compact size, electromagnetic tolerance, chemical inertness, heat resistance, multiplexing capabilities, and suitability for long-distance deployment, FBGs have become essential components in a wide range of applications, such as communication networks, sensing, medical applications, and fibre lasers [3-9]. The growing global demand for precise, reliable, and cost-effective sensing solutions, alongside successful applications across multiple sectors, has driven the widespread adoption of FBGs in recent years.

Traditionally, FBGs are fabricated using phase mask [10, 11] or interference lithography techniques [12, 13] with ultraviolet (UV) exposure [14]. These technologies are highly efficient and deliver outstanding performance, leading to huge commercial success. However, with the increasing demand and diversification of FBG applications, several disadvantages of conventional FBGs fabricated by the aforementioned technologies have become apparent. First, the UV exposure methods generally require either photosensitive optical fibres, such as Ge-doped silica, or hydrogen loading to enhance the RI modulation via photochemical reactions [10]. The fabrication process is complex, and typically involves the removal of the protective coating to allow UV penetration, hydrogen loading to increase photosensitivity, UV exposure to induce RI modulation, hydrogen outgassing to stabilize the structure, and finally recoating to restore mechanical strength [15]. These repeated treatments inevitably degrade the mechanical integrity of the fibre. Moreover, hydrogen diffusion and photochemical instability limit the thermal resistance of such FBGs, leading to gradual degradation at temperatures above 300 °C [16], and complete failure above approximately 450 °C [17] . Consequently, UV-written FBGs are generally unsuitable for high-power fibre-laser applications. Furthermore, due to the inherently weak RI modulation induced by UV exposure, a relatively long grating region is required to achieve high reflectivity. In addition, UV-based methods exhibit poor flexibility and limited material compatibility [18]. Once the phase mask or interference configuration is fixed, only one grating type with a specific period and spectral response can be produced, making parameter adjustment difficult [12, 19]. These methods are also largely ineffective for non-silica fibres—such as fluoride, chalcogenide, or radiation-hardened fibres—which are increasingly important for aerospace, nuclear, and biomedical applications [20-22].

Researchers have been exploring advanced fabrication technologies to overcome the limitations of conventional UV-written FBGs [4, 23, 24]. Among them, the femtosecond (fs) laser inscription technology (FLI) [25, 26] has demonstrated numerous compelling advantages that effectively overcome the limitations of UV-based techniques. By leveraging nonlinear multiphoton absorption, high peak intensity, and ultra-short pulse width [27, 28], fs laser pulses can induce permanent RI changes in a wide variety of optical fibres without requiring photosensitivity enhancement or special doping [29, 30]. These characteristics offers FBG superior thermal and mechanical stability, enabling operation at temperatures exceeding 1000 °C [31]. The FLI technique is compatible with a variety of fibre materials, including standard silica fibres, radiation-hardened fibres, photonic crystal fibres, and even exotic materials such as chalcogenide or sapphire fibres [32-34]. It also can be applied to different fibre types, such as single-mode fibres, multimode fibres, multicore fibres, and coreless fibres [35-37]. This versatility enables the design of complex sensing and laser systems with tailored performance. In addition, with different inscription geometries—point-by-point, line-by-line, or plane-by-plane [20], FLI offers high spatial resolution and remarkable design flexibility, supporting the fabrication of customized grating profiles such as chirped, apodized, and tilted gratings [38-41]. Moreover, the direct, mask-free nature of FLI simplifies fabrication and enables precise in situ writing at arbitrary positions along the fibre, facilitating the realization of multipoint and distributed sensor networks. With the FLI, high-quality FBGs, featuring high reflectivity, low insertion loss, and excellent reliability under harsh environments, such as high pressure, ionizing radiation, and tolerance temperature up to 1900 °C, have been successfully fabricated [4, 29, 42-45]. These unique advantages make FLI a highly versatile and powerful technique for fabricating next-generation FBGs.

However, FLI requires equipment with extremely high precision and stability, as the accurate positioning of the laser focus relative to the fibre core is the most critical factor determining FBG quality. Even slight deviations in focal alignment can lead to irregular RI modulation, spectral distortion, broadened FBG bandwidth, enhanced cladding modes, and increased insertion loss. The process is further complicated by external disturbances such as fibre bending, vibration, temperature fluctuations, and signal noise, which collectively degrade fabrication stability and reproducibility. Consequently, the overall quality of FBGs is strongly depends on the operator's expertise and real-time manual control, resulting in significant batch-to-batch variability. Therefore, precise fibre-core identification and real-time alignment of the laser focus with the core centre represent key challenges in fabricating FBGs with high quality, consistency, and repeatability using FLI, particularly for long FBGs.

Artificial intelligence (AI) has emerged as a powerful tool in advanced manufacturing, especially in image-based process monitoring and quality control [46]. By analysing raw visual data and identifying complex patterns beyond human perception, AI enables highly accurate, real-time decision-making without manual intervention. These capabilities have been widely adopted in industrial inspection, biomedical imaging, and smart manufacturing to improve consistency, reduce human error, and optimise production outcomes [47, 48]. However, its potential in the fabrication of FBGs—particularly for real-time fibre-core detection and laser-focus alignment during FLI—has not yet been fully explored.

In this work, we propose and demonstrate a real-time auto-alignment FLI assisted by AI, which is based on a multiple-layer perceptron (MLP) model. The developed AI-powered FLI can automatically identify fibre cores and dynamically correct the misalignment between the beam focus and the fibre core with high precision and rapid response during high-speed FBG fabrication. Using this AI-powered FLI, we have successfully fabricated four groups of strong FBGs with 8-mm length and weak FBGs with 500-µm length in both AC fibre (SMF-28(R) ULL Fibre) and PI fibre (SM1250(10.4/125) P), demonstrating excellent quality and repeatability. This AI-powered FLI establishes a fully automated, closed-loop fabrication workflow that eliminates the need for manual supervision, ensures consistent fabrication quality, and maintains optimal process stability. By integrating real-time data processing and adaptive feedback control, this AI-powered FLI significantly shortens the fabrication time and offers a promising solution for enhancing the reproducibility, efficiency, and overall reliability of fs laser-based FBG inscription. In addition, this AI-driven system can be extended to fabricate arbitrary FBG structures—from discrete point gratings to complex multi-dimensional patterns—across various fibre types, including multimode, multicore, and polarization-maintaining fibres. With further training and model refinement, the AI-powered FLI provides a scalable and intelligent platform for next-generation automated photonic device manufacturing.

## 2. AI-powered FLI

In this work, the point-by-point strategy of FLI was chosen owing to its simplicity and superior controllability. During FBG fabrication, real-time images captured by the CCD are fed into the trained AI model, which rapidly identifies the precise inscription position of the fibre core, then outputs precise fabrication positions to the 3D electrical stage of the FLI system. As the fabrication position changes, the CCD continuously updates the captured images and send them to the AI model, forming a closed-loop system for real-time FBG fabrication control. A partial schematic of the AI-powered PbP FLI system is shown in Fig. 1a. The system employs a 515 nm *fs* laser with a pulse duration of 250 *fs* and an adjustable repetition rate. The collimated laser beam is tightly focused into the fibre core by a high numerical aperture (0.9) oil-immersion objective with 50x magnification. The fibre is positioned in a V-grooved holder, which is mounted on a 3D electrical stage with nanoscale resolution. A white LED beneath the fibre provides illumination, and the image of the fibre core is directed to a CCD camera via a beam splitter. Fig. 1b presents the flow chart of the AI-powered FBG fabrication. The fibre core recognition AI model is trained using images taken from regions near the fibre core.

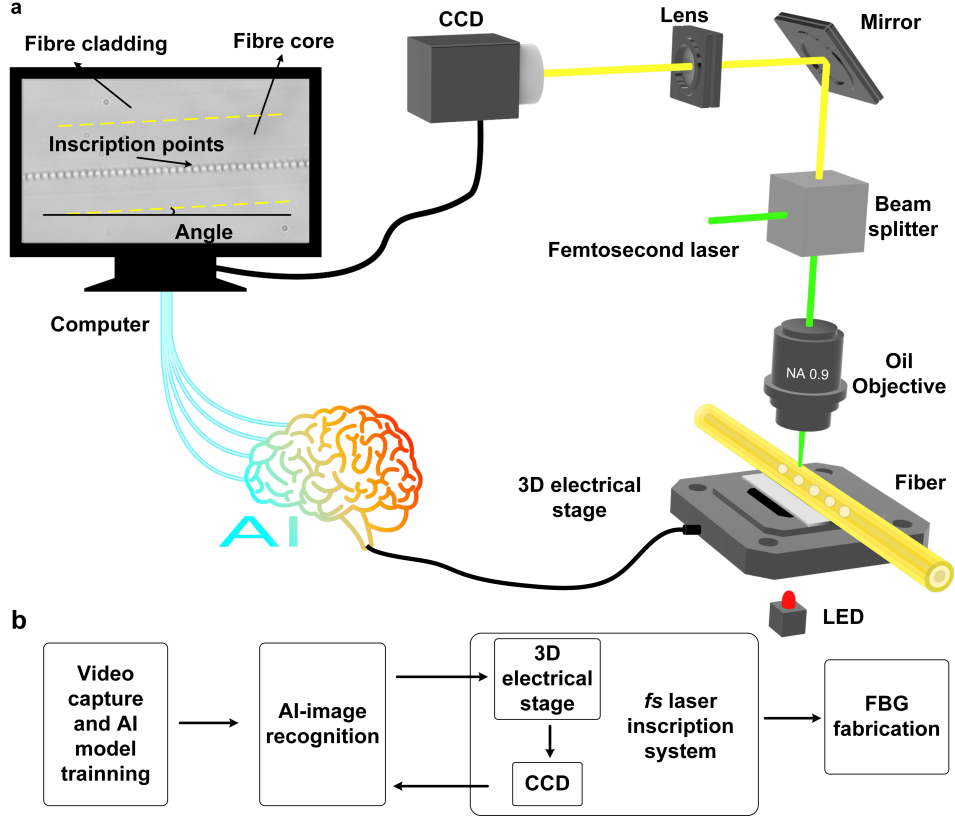

Figure 1. FBG Fabrication by an AI-powered FLI system. (a) Partial schematic of the AI-powered PbP FLI system. (b) flow chart of the AI-powered FBG fabrication process.

The MLP model, known for its mature, stable, and high efficiency is applied to identify the fibre core (Fig. 2b). Each dense layer contains multiple neurons, with the Rectified Linear Unit (ReLU) (equation 1) activation function applied between layers to enhance the model's ability to learn complex feature representations. The input image is first processed by dense layer 1, which includes 32 neurons followed by a ReLU activation. The output then passes through dense layers 2 and 3, each with a different number of neuron and ReLU activation functions, enabling the model to progressively extract high-level features. The final dense layer contains three neurons corresponding to the three image classes (Fig. 2a), followed by a SoftMax function (equation 2) to generate class probabilities.

$$\text{ReLU}: f(x) = \max(0, x) \quad (1)$$

$$\text{SoftMax}: \sigma(z)_i = \frac{e^{z_i}}{\sum_{j=1}^{k} e^{z_i}} \quad (2)$$

where $k$ is the number of classes, and $x$ is the input value before activation. During the fabrication process, the AI model only needs to provide directional guidance to the stage. Specifically, it determines whether the current focal position is too deep (out-of-focus, negative) or too shallow (out-of-focus, positive) relative to the in-focus position, enabling the stage to move in the appropriate direction toward optimal focus. Consequently, the AI model outputs are categorized into three classes: out-of-focus (negative), in-focus, and out-of-focus (positive) (Fig. 2a). This mechanism simplifies the recognition task and improves processing efficiency, making it particularly suitable for real-time FBG inscription systems.

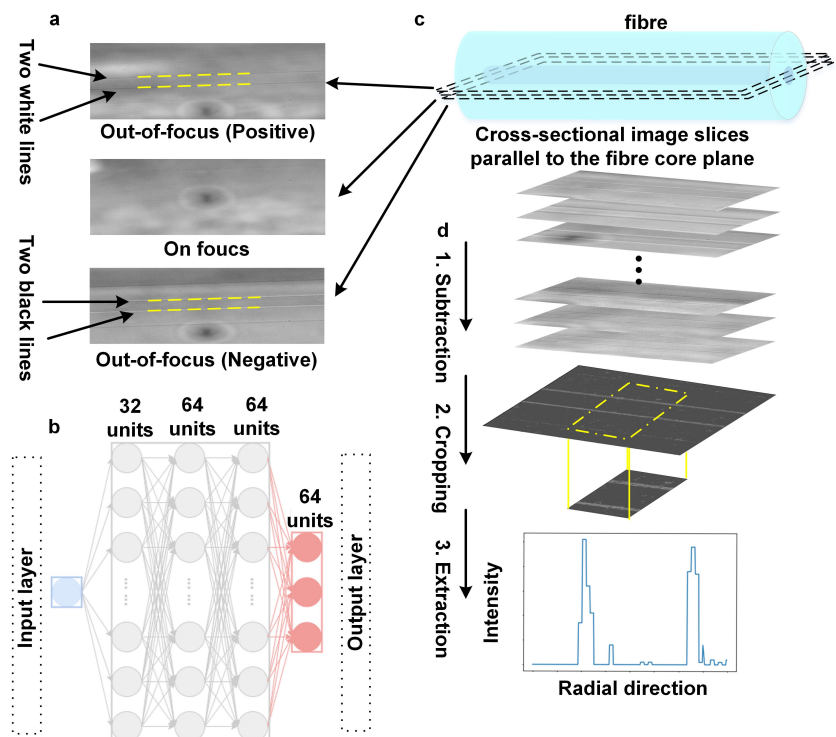

Figure 2. (a) Samples of the out-of-focus (negative), in-focus, and out-of-focus (positive) images of fibre core. (b) The four-dense-layer MLP model, three neurons in the final dense layer. (c) Fibre schematic and the cross-sectional image slices parallel to the fibre core plane. (d) Data preprocessing: subtraction, cropping and extraction.

The training process of the fibre core recognition AI model involves four key stages: data preparation, model designing, training, and evaluation. To prepare the training data, videos segments capturing fibre core, covering focal position out-of-focus (negative), in-focus, and out-of-focus (positive) are collected as the AI training source (Fig. 2c). A total of 45 videos, each with a duration of approximately 45 seconds, were recorded while the stage continuously scanning along the z direction (1,024×1,024 resolution, 23 frame per second) of an AC fibre. Prior to training, two preprocessing steps are applied: frame annotation and de-noising. First, the classification of each image extracted from the videos were manually annotated. Fig. 2a shows examples of the three images classes. When the fibre core is out-of-focus (negative), two white lines appear due to diffraction at the fibre core boundaries. In contrast, the lines become dark when the fibre core is out-of-focus (positive), and the diffraction disappears when the fibre core is in-focus. Those features provide critical cues to identify the focal position. Second, each frame undergoes background subtraction is applied to each frame (Fig. 2d) to enhance the visibility of structural features and reduce noise interference during training. To further improve training efficiency and eliminate irrelevant information, only the region of interest containing the two boundary lines is cropped. The average gray levels along these boundary lines are then calculated and used as the input features for model training.

The MLP model was trained on a dataset of 8,000 images pre-processed from the 45 videos. An early stopping strategy is applied to select the best model based on the loss on validating dataset (2,000 images). The purpose of the training is to minimize the loss through weight updates. The loss function used in the training process is the mean square error (MSE), and the learning rate is set to 0.0001. Equation (3) defines the MSE calculation, and Equation (4) shows the training optimization process.

$$L = \frac{1}{n}\sum_{t=1}^{n}(y_t - y'_t)^2 \qquad (3)$$

$$\theta_l^+ = \theta_l - \tau \times \left(\frac{\partial L}{\partial \theta_l}\right) \qquad (4)$$

where $y_t$ is the true value of a sample and $y'_t$ is the prediction of the sample, $\theta_l$ is the parameter at the $l^{th}$ layer, $\tau$ is the learning rate, $\theta_l^+$ is the updated parameter at the $l^{th}$ layer, and $L$ is the loss. The training process is optimized by monitoring the loss at each step, as illustrated in Fig. 3a. The optimal model is selected at the point where the validation loss reaches its minimum. To further assess the effectiveness of the de-noising process, comparative experiments were conducted under identical conditions (Fig. 3b). The results demonstrate that de-noised images achieve a significantly higher recognition accuracy (>95%) compared to those without de-noising (>65%), highlighting the importance and necessity of this preprocessing step.

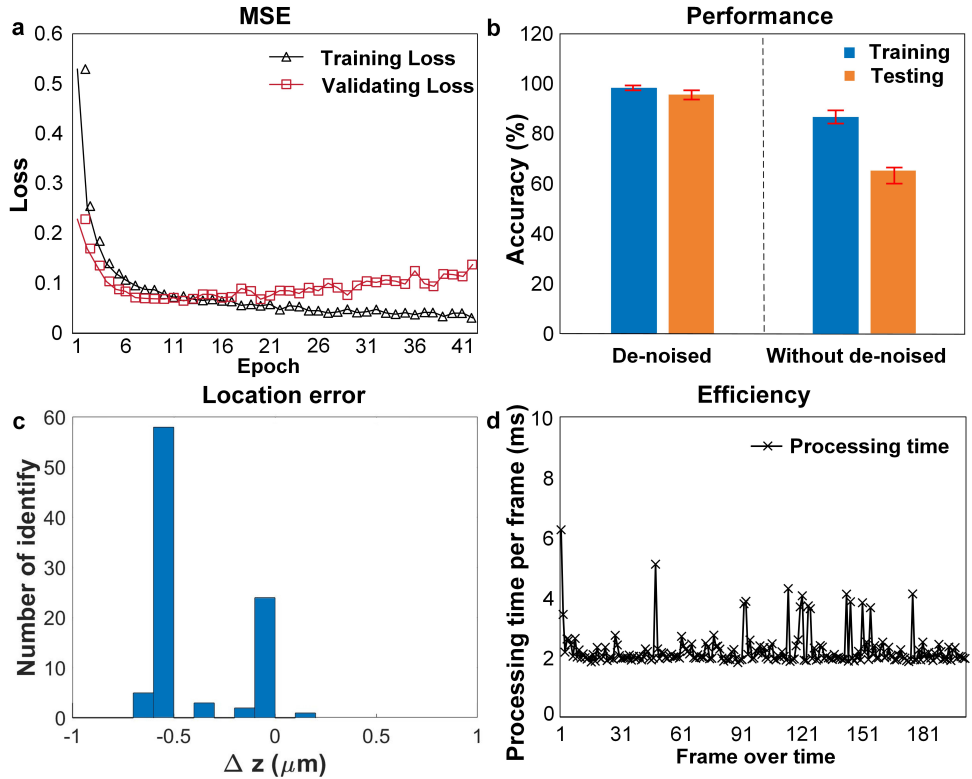

Figure 3. Training and validating assessing of the MLP model for in-focus fibre. (a) Training and validating loss with training epoch number. (b) Accuracy comparisons of the denoised and without de-noised data processing. (c) Fibre core recognition number and location error with repeated measurements. (d) Average prediction time per image for fibre core recognition.

Accuracy and efficiency are critical metrics for evaluating the performance of an AI model. The localization accuracy is assessed by conducting 100 repeated tests to identify the in-focus position of the fibre core. The resulting position errors ranged from –0.6 μm to 0.2 μm, demonstrating the model's nanometer-scale precision. Notably, a consistent offset toward the negative direction was observed, indicating that introducing a compensation correction could further enhance the model's accuracy. The fibre core recognition efficiency is evaluated based on the image inference time, as shown in Fig. 3d. The average prediction time per image is approximately 0.002 seconds (measured on a Windows laptop equipped with an Intel Core i7-10875H @ 2.30 GHz CPU, 32 GB RAM, and RTX 3070 GPU), corresponding to a frame rate of around 500 FPS. This rapid inference speed makes the model well-suited for real-time fibre core recognition tasks.

## 3. Experimental results of AI-powered FBG fabrication

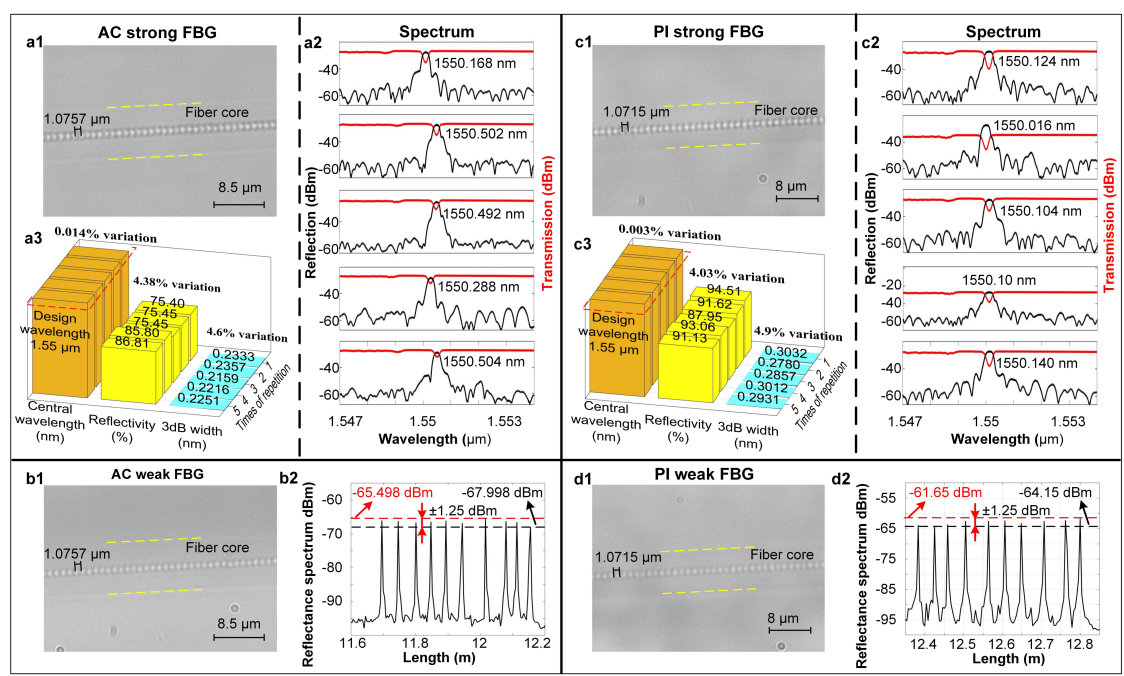

Figure 4. FBG fabrication using this AI-powered FLI system. (a1, b1, c1, and d1) Microscopic images of the fabricated FBs. (a2 and c2) Measured spectrums of the five strong FBGs using AC and PI fibres. (a3 and c3) Corresponding statistical analysis of the central wavelength, reflectivity and 3dB width of the strong FBGS. (b2 and d2) Spectrum of the ten weak FBGs using AC and PI fibres.

To verify the stability and repeatability of the AI-powered FLI system in FBG inscription, we designed an 8-mm-long strong FBG and a 500-μm-long weak FBG, both with a central wavelength of 1550 nm. Using AC fibres, five strong FBGs and ten weak FBGs were inscribed. The fabrication and structural parameters of the FBGs are listed in Table 1. The fabrication process begins with an automatic, precise alignment procedure to ensure optimal positioning and laser focusing before inscription. A video of the entire automatic fabrication process is provided in the Supplementary Materials. Figs. 4a1 and 4b1 show partial microscopic images of the strong and weak FBGs inscribed in AC fibres. The period of the RI modulation points is 1.0757 μm. Notably, the strong FBGs exhibits a controlled apodization profile, with the RI modulation lines tilted by a slight 0.002-rad relative to the fibre core boundary, ensuring smooth amplitude variation. The fs laser power used for inscription is 2.1 mW, and the fabrication speed is 0.5 mm/s. The reflection spectra of the strong and weak FBGs were measured using an AQ6370D Optical Spectrum Analyzer and a LUNA Spectrum Analyzer with a fibre laser source (ANDO AQ4321D, operating in the 1520–1620 nm range), respectively. Fig. 4a2 is the measured reflection spectra of the five strong FBGs, showing central wavelengths of 1550.391 ± 0.223 nm under identical fabrication parameters, resulting in a wavelength variation of less than 0.014%. Statistical analysis is presented in Fig. 4a3. The 3dB width is 0.2263 ± 0.0104 nm, with a variation of less than 4.6%, and the reflectivity is approximately 79.78 ± 4.38%, corresponding to a variation of less than 5.5%. Fig. 4b2 shows the reflection spectra of the ten weak FBGs fabricated under the same conditions. The measured reflected power is –66.43 ± 1.25 dBm, with a variation of less than 1.88%. Taking into account unavoidable measurement errors, the statistical results demonstrate the excellent stability and repeatability of the AI-powered FLI system in fabricating high-quality FBGs.

Table 1. Fabrication and structure parameters of the testing FBGs

| Fibre type | FBG types | Period (μm) | Length (μm) | Laser power (mW) | Fabrication speed (mm/s) | Tilt angle (rad) |
|---|---|---|---|---|---|---|
| AC | Strong | 1.0757 | 5000 | 2.1 | 0.5 | 0.002 |
| AC | Weak | 1.0757 | 500 | 2.1 | 0.5 | 0 |
| PI | Strong | 1.0715 | 5000 | 2.3 | 0.5 | 0.002 |
| PI | Weak | 1.0715 | 500 | 2.1 | 0.5 | 0 |

Moreover, to further evaluate the feasibility and stability of the AI-powered FBG inscription system, PI fibres was used to fabricate five strong and ten weak FBGs. The fabrication and structural parameters of the PI-based FBGs were optimized to meet the design requirements and summarized in Table 1. Figs. 4c1 and 4d1 present partial microscopic images of the strong and weak FBGs inscribed in PI fibres. The period of the RI modulation points is 1.0715 μm. For the strong FBGs, the RI modulation line exhibits a 0.002-rad angle with respect to the fibre core boundary. The fs laser power used is 2.3 mW for the strong FBGs and 2.1 mW for the weak FBGs, with a consistent fabrication velocity of 0.5 mm/s. Fig. 4c2 shows the reflection spectra of the five strong FBGs inscribed in PI fibres. The measured central wavelength is 1550.097 ± 0.043 nm, corresponding to a variation of less than 0.003%. Fig. 4c3 presents the statistical analysis: the 3dB width is 0.2922 ± 0.0142 nm, with a variation of less than 4.9%, and the reflectivity is 91.65 ± 3.70%, with a variation of less than 4.03%. Fig. 4d2 displays the reflection spectra of the ten weak FBGs fabricated under the same conditions. The measured reflected power is –62.85 ± 1.25 dBm, with a variation of less than 2.0%. Considering the negligible measurement errors, these results further demonstrate the feasibility, scalability and stability of the AI-powered FLI system in fabricating high-quality FBGs. Therefore, the proposed AI-powered FLI system shows strong potential to become one of the most efficient and user-friendly techniques for large-scale FBG fabrication.

## 4. Conclusions

In conclusion, we have developed an AI-powered FLI system for the automatic fabrication of FBGs. The AI model based on the MLP only provides directional guidance to the motion control stage, significantly simplifying the recognition task and enhancing processing efficiency. This design makes the system particularly suitable for real-time FBG inscription. Using this system, ten strong FBGs (8 mm in length) and twenty weak FBGs (500 μm in length) were successfully fabricated in two types of optical fibres—AC and PI fibres. The critical parameters of both the strong and weak FBGs were characterized. The minimal variations in these measured parameters demonstrate the high quality, repeatability, and stability of the developed AI-powered FLI system. As a result, the proposed AI-powered FLI system shows great potential for high-efficiency and high-throughput fabrication of FBGs. Furthermore, it holds promise for extension to the fabrication of diverse types of FBGs, including strong and weak FBGs, point-by-point, line-by-line, plane-by-plane, and tilted structures, across various fiber types such as single-mode, multimode, multicore, and polarization-maintaining fibers.

**Funding.** This research was funded by the Australian Research Council (ARC) through the Discovery Project scheme (Grant Nos. DP190103186, DP220100603), the ARC Industrial Transformation Research Hubs (Grant No. IH240100009), the ARC Centre of Excellence Program (Grant No. CE230100006), the ARC Future Fellowship scheme (Grant No. FT210100806), the ARC Linkage Program (Grant No. LP210200345), the ARC Future Fellowship scheme (Grant No. FT220100559), the ARC Linkage Program (Grant No. LP210100467)

**Disclosures.** The authors declare that there are no conflicts of interest related to this article.

**Data availability.** Data underlying the results presented in this paper may be obtained from the authors upon reasonable request.

**Author Contributions.** Wenbo Liu and Guiyuan Cao contributed equally.